\RequirePackage{lineno}
\documentclass[%
 reprint,
superscriptaddress,
 amsmath,amssymb,
 aps,
]{revtex4-2}

\usepackage{graphicx}   
\usepackage{placeins}
\usepackage{xcolor}
\usepackage{natbib}
\usepackage{hyperref}
\usepackage{amsmath,amssymb,amsfonts}
\usepackage{xspace}
\usepackage{subfigure}
\usepackage{soul}
\usepackage{booktabs} 
\usepackage{graphicx}
\usepackage{epstopdf}
\usepackage{comment}

\newcommand{\five}        {$\sqrt{s_{\mathrm{NN}}}~=~5.36$~TeV\xspace}

\newcommand{\ppt}{\ensuremath{p_{\rm T}}\xspace}

\newcommand {\auau}  {\ensuremath{\rm Au+Au}\xspace}
\newcommand {\pbpb}  {\ensuremath{\rm Pb+Pb}\xspace}

\newcommand {\oo}  {\ensuremath{\rm O+O}\xspace}
\newcommand {\nene}  {\ensuremath{\rm Ne+Ne}\xspace}
\newcommand {\pp}    {\ensuremath{\rm p+p}\xspace}

\newcommand {\pA}    {\ensuremath{\rm p+A}\xspace}

\setstcolor{red}

\usepackage[bottom]{footmisc}

\begin{document}

\title{Impact of nuclear deformation on particle production in \nene collisions at \texorpdfstring{\five}{sqrt(sNN)=5.36 TeV} from AMPT-SM }

\author{M.~U.~Ashraf}
\email{muashraf@wayne.edu}
\affiliation{Department of Physics and Astronomy, Wayne State University, 666 W. Hancock, Detroit, Michigan 48201, USA}%

\author{A.~M.~Khan}
\email{akhan129@gsu.edu}
\affiliation{Department of Physics and Astronomy, Georgia State University, Atlanta, GA, 30303, USA}%

\author{M.~Shahid}
\email{mshahid1@gsu.edu}
\affiliation{Department of Physics and Astronomy, Georgia State University, Atlanta, GA, 30303, USA}%

\author{Faraz Mohd Mehdi}
\email{farazmmehdi@wayne.edu}
\affiliation{Department of Physics and Astronomy, Wayne State University, 666 W. Hancock, Detroit, Michigan 48201, USA}%

\date{\today}

\begin{abstract}

We present a systematic study of particle production in \nene collisions at \five using the A Multi-Phase Transport (AMPT) model with string melting (SM) configuration. The analysis compares spherical and deformed configurations of ${}^{20}\mathrm{Ne}$ to investigate the influence of initial-state nuclear deformation on bulk observables. Charged-particle pseudorapidity ($\langle dN_{\mathrm{ch}}/d\eta \rangle$) densities, identified particle yields ($dN/dy$), transverse momentum (\ppt) spectra, mean transverse momentum ($\langle p_{\mathrm{T}} \rangle$), and $p_{\mathrm{T}}$-differential particle ratios ($K/\pi$ and $p/\pi$) are studied as functions of multiplicity and centrality. The results show that all observables exhibit the expected dependence on event activity, including smooth multiplicity scaling, mass ordering in $\langle p_{\mathrm{T}} \rangle$, and characteristic features associated with radial flow and quark coalescence. Differences between the two configurations on bulk observables remain small across all observables, typically at the level of a 2\%--6\% percent, with slightly enhanced sensitivity observed in peripheral collisions. These findings suggest that, within the AMPT-SM framework, the collective dynamics and hadrochemical composition are primarily governed by the overall system density and interaction dynamics, while the influence of initial-state deformation is subleading. This study provides a baseline for understanding deformation effects in light-ion collision systems and highlights the limited sensitivity of bulk observables to initial nuclear geometry in transport-based approaches.

\end{abstract}

\maketitle

\section{Introduction}
\label{sec1}

Over the past decade, ultra-relativistic heavy-ion collisions have established the formation of a strongly interacting quark–gluon plasma (QGP) in large systems such as \pbpb collisions at the Large Hadron Collider (LHC) and \auau collisions at Relativistic Heavy-Ion Collider (RHIC). This medium exhibits collective behavior consistent with nearly ideal hydrodynamic expansion, as reflected in anisotropic flow, long-range correlations, and characteristic patterns in identified particle production \cite{Heinz2013,Shuryak2017}.

In recent years, similar QGP-like signatures have been observed in smaller collision systems, including proton–proton (\pp) and proton–nucleus (\pA) collisions~\cite{ALICE2017strangeness}. Measurements from the various LHC experiments have revealed long-range ridge structures, collective flow patterns, and multiplicity-dependent strangeness enhancement \cite{Khachatryan2010,CMS2013ridge,ALICE2017strangeness,ATLAS2016flow}. These observations have motivated systematic investigations of the role of system size and initial conditions in the development of collective phenomena \cite{Busza2018,Nagle2018}.

Light-ion collisions, such as oxygen–oxygen (\oo) and neon–neon (\nene), provide an intermediate system size that enables controlled studies of the transition between small and large collision systems. By varying the system size while maintaining high collision energies, these systems offer a framework to probe the onset of collective behavior and to examine the influence of initial-state geometry on final-state observables \cite{Loizides2016,ALICEOO2022}.

In this context, nuclear deformation introduces an additional structure in the initial-state geometry through deviations from spherical symmetry. In this work, we consider both a spherical configuration and a deformed configuration of ${}^{20}\mathrm{Ne}$ nuclei within a Woods–Saxon (WS) density profile. The deformed case includes non-zero quadrupole and octupole deformation parameters, leading to modified spatial distributions of nucleons and, consequently, changes in the initial overlap geometry. This provides a controlled setup to study the sensitivity of bulk particle production to initial-state deformation effects.

To investigate these effects, we employed A Multi-Phase Transport model with string melting configuration (AMPT-SM)~\cite{Lin2005}. The AMPT-SM framework incorporates partonic scatterings followed by hadronization via quark coalescence and subsequent hadronic interactions, allowing for a consistent description of the system evolution from initial conditions to final-state particles.

The AMPT-SM has been widely used to describe particle production and collective behavior across a broad range of collision systems and energies. It has been shown to provide a reasonable description of transverse momentum spectra, particle yields, and identified particle ratios in both large and small systems, including \auau and \pbpb, \oo as well as \pA and high-multiplicity \pp collisions~\cite{Lin2005,Khan:2024fef,Ashraf:2024ocb,Bashir:2025epjp,Xu2011}. In particular, AMPT-SM has been widely used to study bulk observables such as $\langle p_{\mathrm{T}} \rangle$ and particle ratios, providing a useful framework to investigate the interplay between partonic interactions and hadronization via quark coalescence\cite{Lin2005,Khan:2024fef,Ashraf:2024ocb,Bashir:2025epjp,Xu2011}. This provides a transport-based framework to study the influence of initial-state geometry, including nuclear deformation, on final-state particle production.

The present analysis focuses on identified particle observables in \nene collisions, including charged-particle pseudorapidity ($\langle dN_{\mathrm{ch}}/d\eta \rangle$) densities, transverse momentum (\ppt) spectra, rapidity density ($dN/dy$), mean transverse momentum ($\langle p_{\mathrm{T}} \rangle$), and \ppt-differential particle ratios such as $K/\pi$ and $p/\pi$. These observables probe different aspects of the collision dynamics, with \ppt spectra and $\langle p_{\mathrm{T}} \rangle$ reflecting collective expansion, while particle ratios provide insight into hadrochemistry and particle production mechanisms. By comparing spherical and deformed configurations, this study examines the extent to which initial-state deformation influences bulk particle production in intermediate-size collision systems.

\section{Event Generation and Analysis Methodology}
\label{sec2}

This section provides an overview of AMPT-SM model used in the study, followed by a description of the nuclear density distributions for $^{20}\mathrm{Ne}$.

\subsection{A Multi-Phase Transport (AMPT) model and nuclear structure}

AMPT is a multiphase transport framework extensively used to study relativistic nuclear collisions~\cite{Lin:2004en, Lin2005}. It has four well-defined components that link the initial nuclear geometry to final-state hadron production. The initial conditions are provided by the HIJING~\cite{Wang:1991hta} model, which generates the spatial distribution of participant nucleons and produces excited strings from hard and soft interactions. A relativistic transport (ART) model~\cite{Li:1995pra} is used for the final-state hadronic interactions. The dynamical evolution of the partonic stage is described via Zhang’s Parton Cascade (ZPC)~\cite{Zhang:1997ej}, including two-body elastic collisions with different cross section:
\begin{equation}
    \frac{d\sigma}{dt} \approx \frac{9\pi \alpha_s^2}{2(t-\mu^2)^2},
\end{equation}
where $t$ denotes the Mandelstam variable corresponding to the four-momentum transfer. In this study, the strong coupling constant is set to $\alpha_{s}=0.33$ and the parton screening mass to $\mu=3.2~\mathrm{fm}^{-1}$, corresponding to a parton scattering cross section of $1.5~\mathrm{mb}$~\cite{Xu:2011fi, Xu:2011fe}. Hadrons are formed through quark coalescence together with Lund string fragmentation. In AMPT-SM, all excited strings are converted into their constituent valence quarks and antiquarks. The momentum distribution of these partons follows the Lund string fragmentation function:
\begin{equation}
f(z) \propto z^{-1} (1 - z)^{a} \exp\left(-\frac{b\,m_{T}^{2}}{z}\right),
\end{equation}
where \(z\) is the light-cone momentum fraction carried by the produced hadron, \(m_{T}\) is the transverse mass, and \(a\) and \(b\) are the Lund fragmentation parameters. In this work, we set $a = 0.5$ and $b = 0.9$ $\mathrm{GeV}^{-2}$ motivated from the earlier studies~\cite{Xu:2011fi, Xu:2011fe, Wang:2025mmi}. The nucleon density distribution in AMPT-SM is described by a WS distribution: 
\begin{equation}
\rho(r,\theta) = \frac{\rho_{0}(1+\omega(\frac{r}{R_0})^2)}{1 + \exp\left[(r - R(\theta -\phi))/d\right]},
\end{equation}
where $\rho_0$ denotes the normal nuclear density, $r$ is the distance from the center of the nucleus, $d$ is the surface diffuseness. For deformed nuclei, the nuclear surface is parameterized by an $\theta, \phi$ radius defined as:
\begin{equation}
R(\theta -\phi) = R_{0}\left[1 + \beta_{2} Y_{20}(\theta) + \beta_{3} Y_{30}(\theta)\right],
\end{equation}
where $R_0$ is the radius parameter, \(\beta_{2}\) and \(\beta_{3}\) denote the quadrupole and octupole deformation parameters, respectively, and \(Y_{20}\) and \(Y_{30}\) are spherical harmonics.

\begin{table}[t]
\centering
\caption{Woods--Saxon (WS) parameters used in the AMPT-SM for \({}^{20}\mathrm{Ne}\), where both spherical and deformed configurations are considered.}
\renewcommand{\arraystretch}{1.25}
\begin{tabular}{lcccc}
\toprule
Nucleus & $R_{0}$ (fm) & $d$ (fm) & $\beta_{2}$ & $\beta_{3}$ \\
\midrule
${}^{20}\mathrm{Ne}$ & 2.256 & 0.594 & 0     & 0     \\
${}^{20}\mathrm{Ne}$  & 2.256 & 0.594 & 0.548 & 0.281 \\
\bottomrule
\end{tabular}\label{tab1}
\end{table}

\begin{figure}
    \centering
    \includegraphics[width=\linewidth]{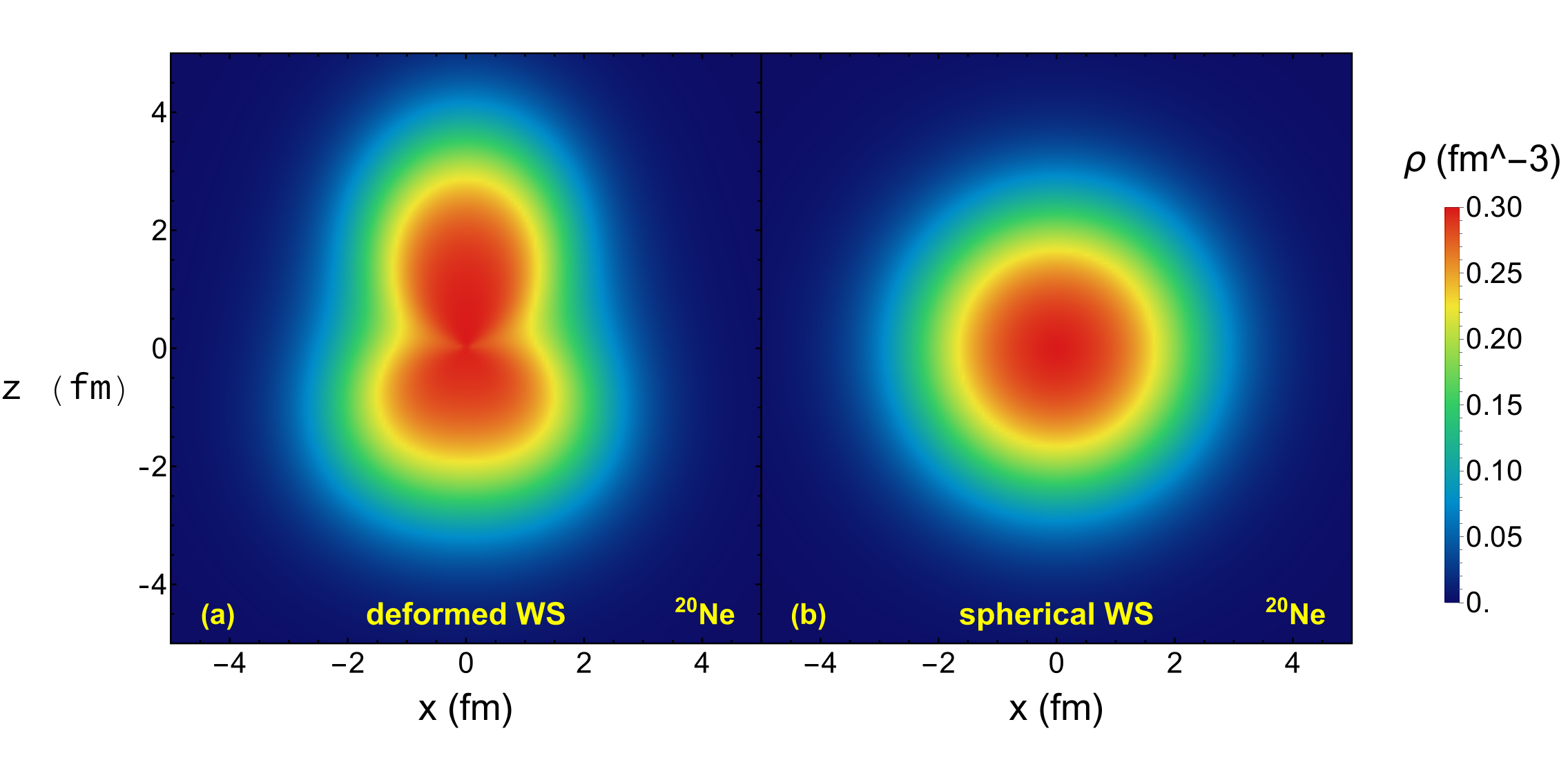}
    \caption{(Color Online) Density contours of \({}^{20}\mathrm{Ne}\) with a deformed WS distribution (left), and \({}^{20}\mathrm{Ne}\) with a spherical WS distribution (right).
}
    \label{fig1}
\end{figure}

To describe non-equilibrium dynamics in light-ions collisions, transport models provide a natural framework since they do not rely on the assumption of local thermal equilibrium. In this study, events are generated using two configurations for \({}^{20}\mathrm{Ne}\) spherical and deformed. The corresponding parameter values are motivated by Ref.~\cite{Wang:2025mmi} and are summarized in Table~\ref{tab1}. The density contours of \({}^{20}\mathrm{Ne}\) for different configurations are shown in Fig.~\ref{fig1}. A total of 10 million minimum bias events have been analyzed for each case. Once experimental data for the corresponding systems become available, the above parameters will be further constrained and refined, which is beyond the scope of the present study.

\section{Results and Discussion}
This section presents a quantitative comparison of particle production in \nene collisions at \five for default and deformed nuclear configurations. The analysis focuses on the sensitivity of bulk observables to initial-state geometry.

\begin{figure}[h]
\centering
\includegraphics[width=\columnwidth]{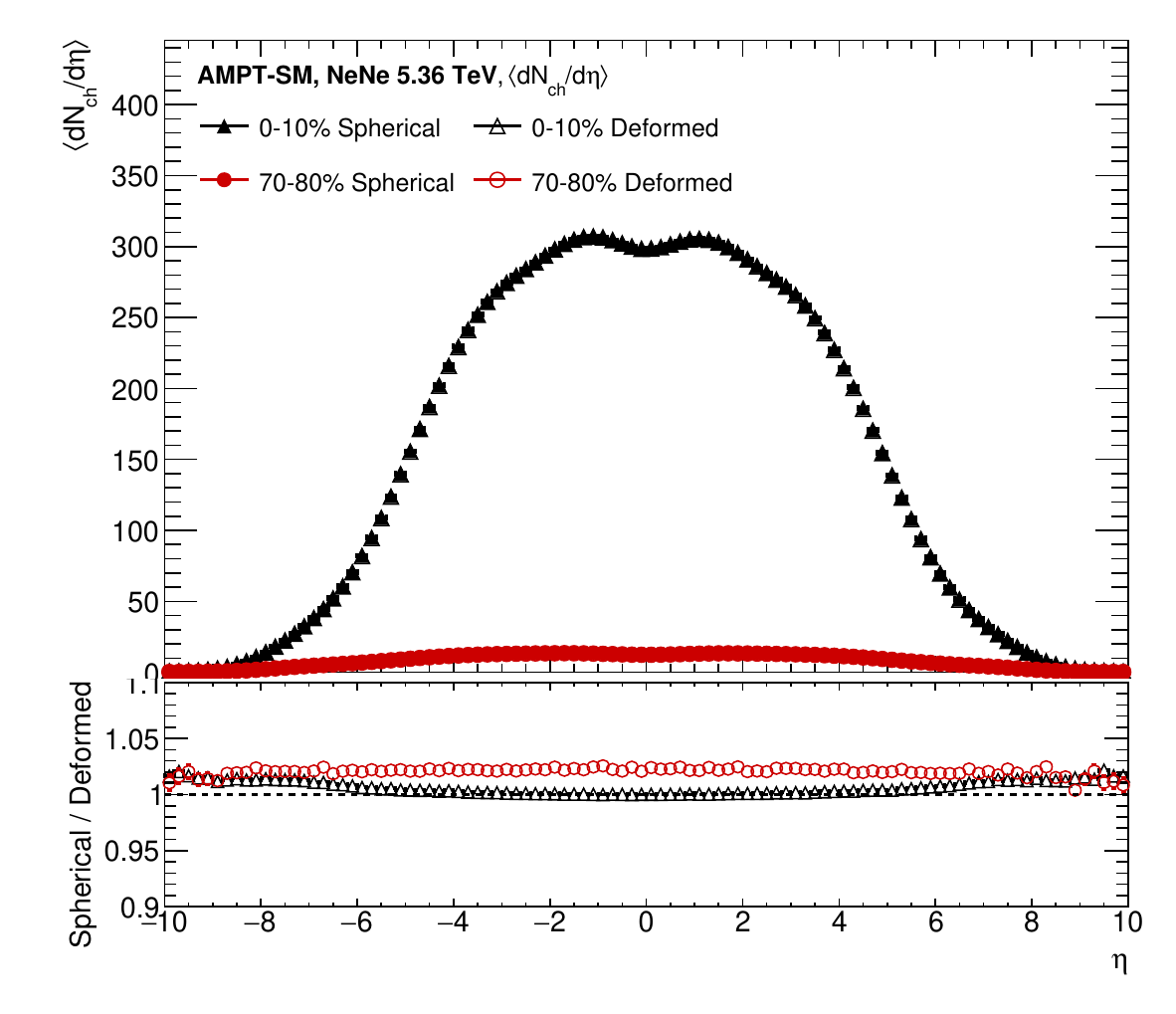}
\caption{(Color Online) charged particle multiplicity ($\langle dN_{\mathrm{ch}}/d\eta \rangle$) distributions at $|\eta| < 0.8$ in 0--10\% central and 70--80\% peripheral \nene collisions at \five from AMPT-SM. Results for spherical and deformed configurations of ${}^{20}\mathrm{Ne}$ are shown. The lower panel presents the ratio of spherical to deformed results. Statistical uncertainties are smaller than the marker size.}
\label{fig2}
\end{figure}

Figure~\ref{fig2} shows the charged-particle multiplicity ($\langle dN_{\mathrm{ch}}/d\eta \rangle$), at mid-rapidity ($|\eta| < 0.8$) for 0--10\% central and 70--80\% peripheral \nene collisions at \five from AMPT-SM. The results are presented for both spherical and deformed configurations of ${}^{20}\mathrm{Ne}$. In central collisions, the $\langle dN_{\mathrm{ch}}/d\eta \rangle$ distributions for the two configurations are nearly same, indicating that nuclear deformation has a negligible effect on the overall particle production in high-multiplicity events. This behavior suggests that, in central collisions, the large number of participant nucleons and strong interaction dynamics dominate the particle production, effectively washing out differences arising from the initial nuclear geometry. In contrast, a small but noticeable difference is observed in peripheral collisions, where the effect of nuclear deformation reaches approximately $\sim 4\%$, which can be seen in the ratio pannel at the bottom. In this regime, the particle density is lower and the number of interactions is reduced, allowing a modest imprint of the initial-state geometry to persist in the final-state particle production. Nevertheless, the overall effect remains small, indicating that bulk particle production is largely governed by event activity rather than detailed geometric features of the colliding nuclei.

\begin{figure}[h]
\centering
\includegraphics[width=\columnwidth]{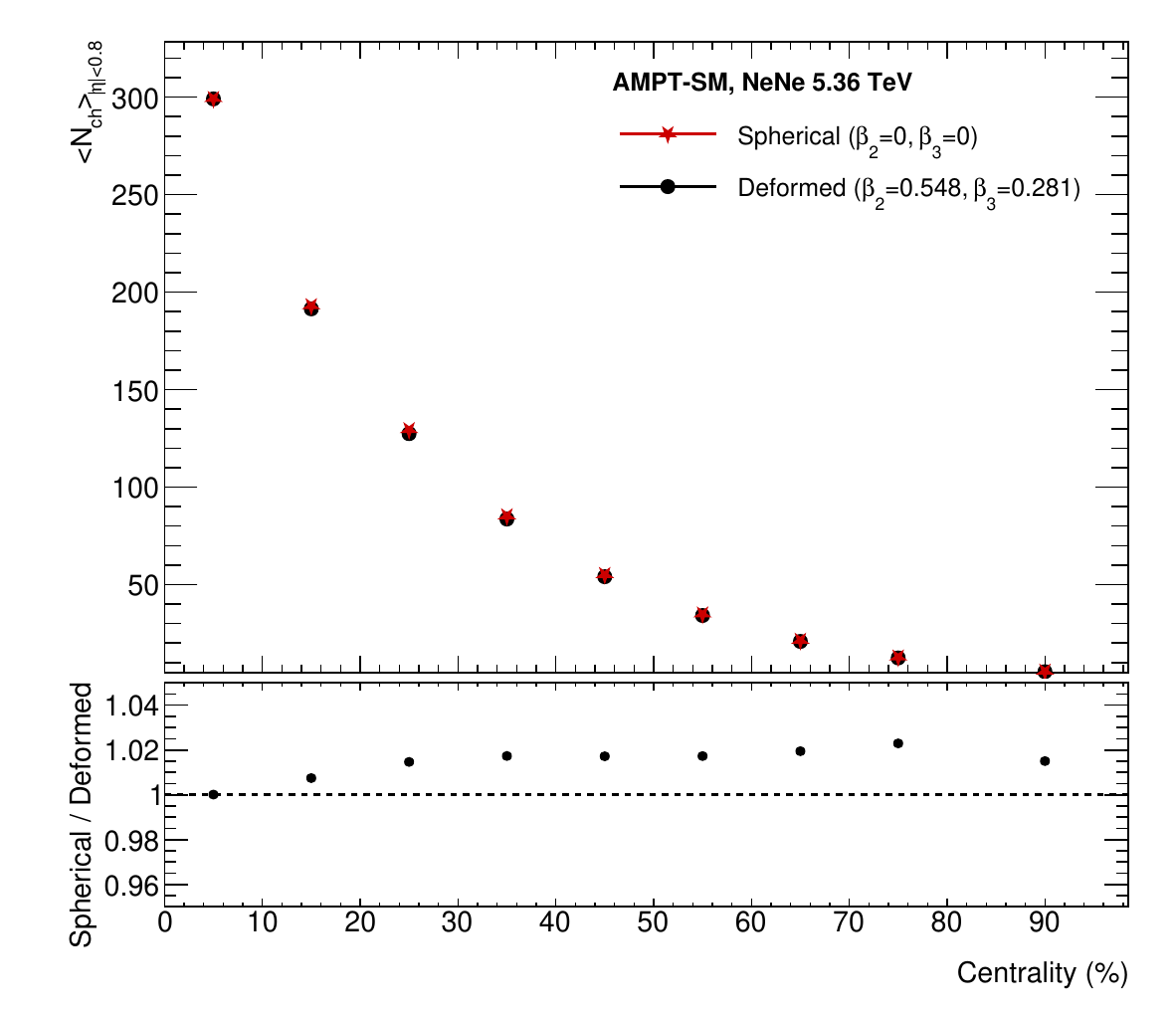}
\caption{(Color Online) Average charged-particle multiplicity at mid-rapidity ($|\eta| < 0.8$) as a function of centrality in \nene collisions at \five from the AMPT-SM. Results for spherical and deformed configurations of ${}^{20}\mathrm{Ne}$ are shown. The lower panel presents the ratio of spherical to deformed results. Statistical uncertainties are smaller than the marker size.}
\label{fig2a}
\end{figure}

Figure~\ref{fig2a} shows the average charged-particle multiplicity at mid-rapidity ($|\eta| < 0.8$) as a function of centrality for \nene collisions at \five, obtained using the AMPT-SM. The charged-particle yield decreases smoothly from central to peripheral collisions, reflecting the expected reduction in particle production with decreasing system size and the corresponding decrease in the number of participating nucleons. A comparison between spherical and deformed configurations of ${}^{20}\mathrm{Ne}$ indicates that the overall behavior of the charged-particle yield is very similar across all centralities. The ratio of the two configurations remains close to unity, with deviations at the level of a $\approx$ 2 percent. A slightly larger deviation from unity is observed towards peripheral collisions, where the particle density is lower and fewer interactions occur, suggesting a marginal increase in sensitivity to nuclear deformation. This suggests that the inclusion of nuclear deformation, characterized by non-zero quadrupole and octupole deformation parameters, has a limited impact on global particle production at mid-rapidity within the AMPT-SM framework. Since centrality classes are defined using multiplicity-based quantiles of the RefMult distribution, the observed trends are largely driven by the event classification procedure. Therefore, the small differences between spherical and deformed configurations indicate that deformation does not introduce significant modifications beyond the centrality selection effects in this observable.

\begin{figure}[h]
\centering
\includegraphics[width=\columnwidth]{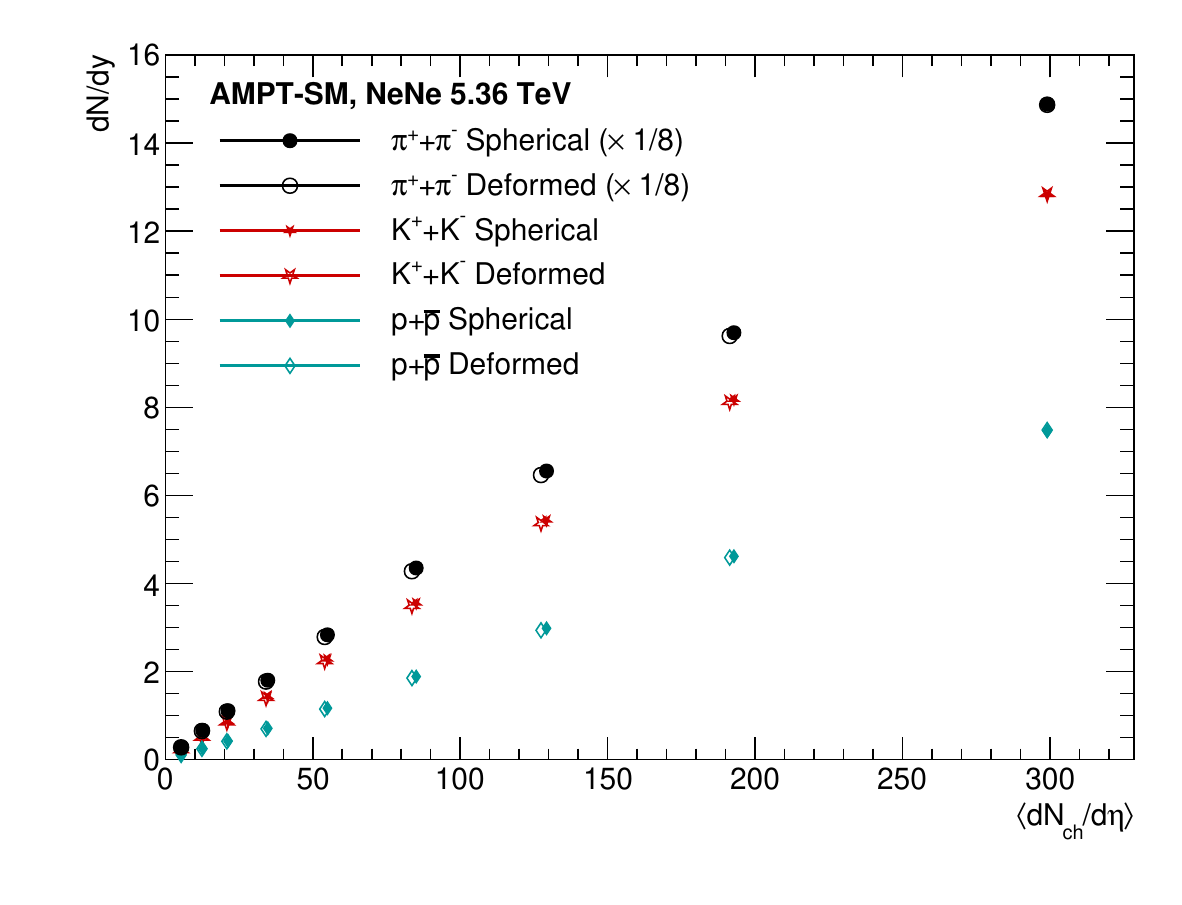}
\caption{(Color Online) Rapidity density $dN/dy$, of identified charged hadrons as a function of the average charged-particle multiplicity $\langle dN_{\mathrm{ch}}/d\eta \rangle$, in \nene collisions at \five from the AMPT-SM. Results are shown for $\pi^{+}+\pi^{-}$, $K^{+}+K^{-}$, and $p+\bar{p}$ for spherical and deformed configurations of ${}^{20}\mathrm{Ne}$. For visual clarity, the pion yields are scaled by a factor of $1/8$.}
\label{fig3}
\end{figure}

Figure~\ref{fig3} presents the rapidity density $dN/dy$, of identified hadrons as a function of the average charged-particle multiplicity $\langle dN_{\mathrm{ch}}/d\eta \rangle$. The yields of $\pi^{+}+\pi^{-}$, $K^{+}+K^{-}$, and $p+\bar{p}$ increase smoothly with multiplicity, reflecting the dominant role of event activity in particle production. This behavior is consistent with expectations from the AMPT-SM framework, where the number of produced partons and their subsequent hadronization scale with the system size. A clear hierarchy in particle yields is observed, with $\pi^{+}+\pi^{-}$ production dominating over $K^{+}+K^{-}$ and $p+\bar{p}$. This is primarily due to the lower mass of pions, which makes their production energetically favorable in the soft particle production regime and is consistent with the predictions of thermalized Boltzmann production of secondary particles in high-energy nuclear collisions.

At fixed multiplicity, the results for spherical and deformed configurations remain very similar for all particle species, indicating that nuclear deformation has only a limited impact on the integrated production of identified hadrons. This suggests that, within the AMPT-SM model, the effects of initial-state geometry are largely diluted during the partonic and hadronic evolution of the system. Consequently, bulk particle yields are primarily governed by the overall multiplicity rather than the detailed shape of the initial nuclear density distribution. No significant species-dependent deviation is observed between spherical and deformed configurations, implying that deformation does not strongly affect hadrochemical composition at the level of integrated yields. Any potential deformation-induced effects appear to be subleading compared to the multiplicity-driven scaling of particle production.

\begin{figure*}[t]
\centering

\begin{minipage}[t]{0.33\textwidth}
  \centering
  \includegraphics[width=\linewidth]
  {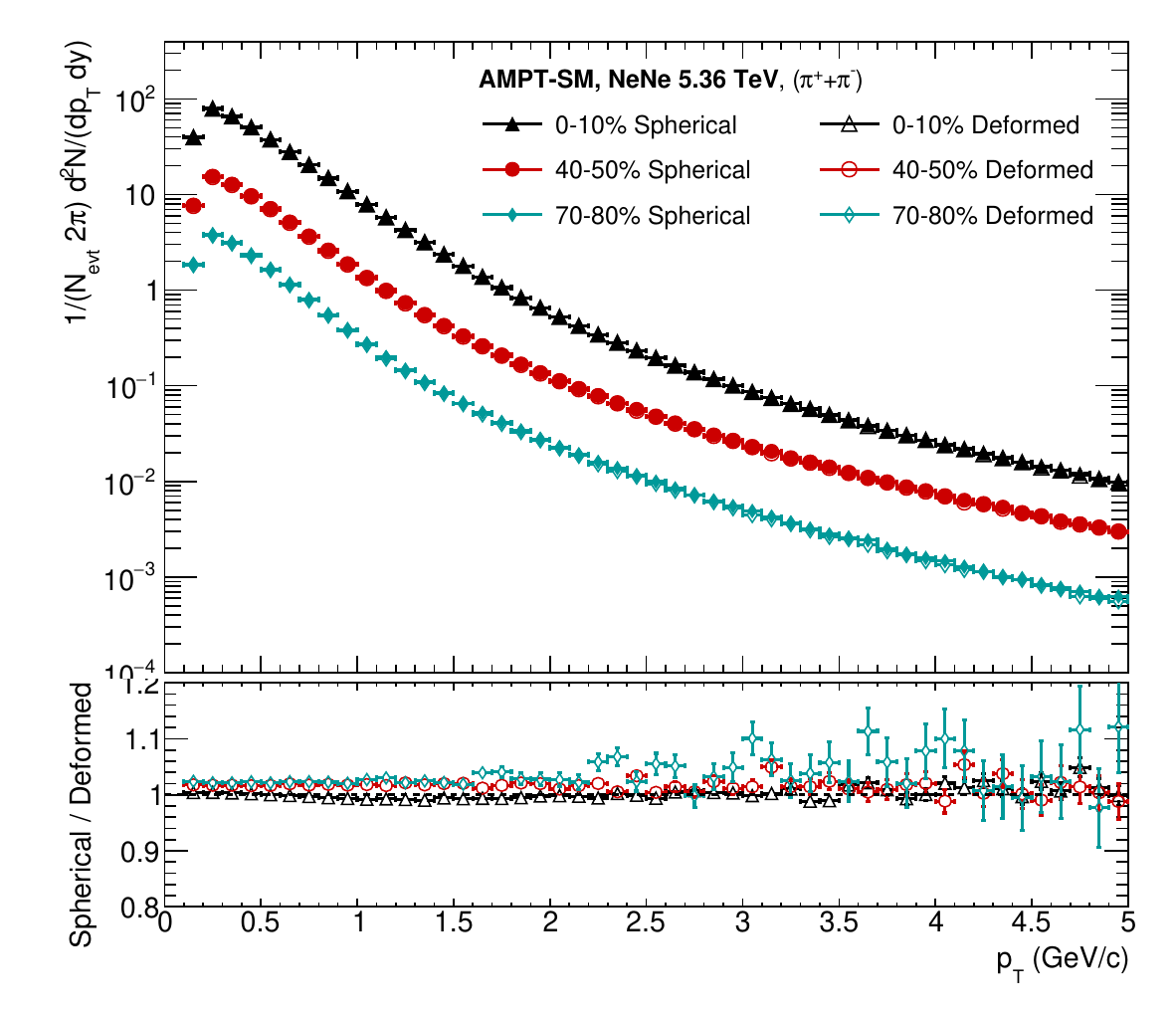}
\end{minipage}\hfill%
\begin{minipage}[t]{0.33\textwidth}
  \centering
  \includegraphics[width=\linewidth]
  {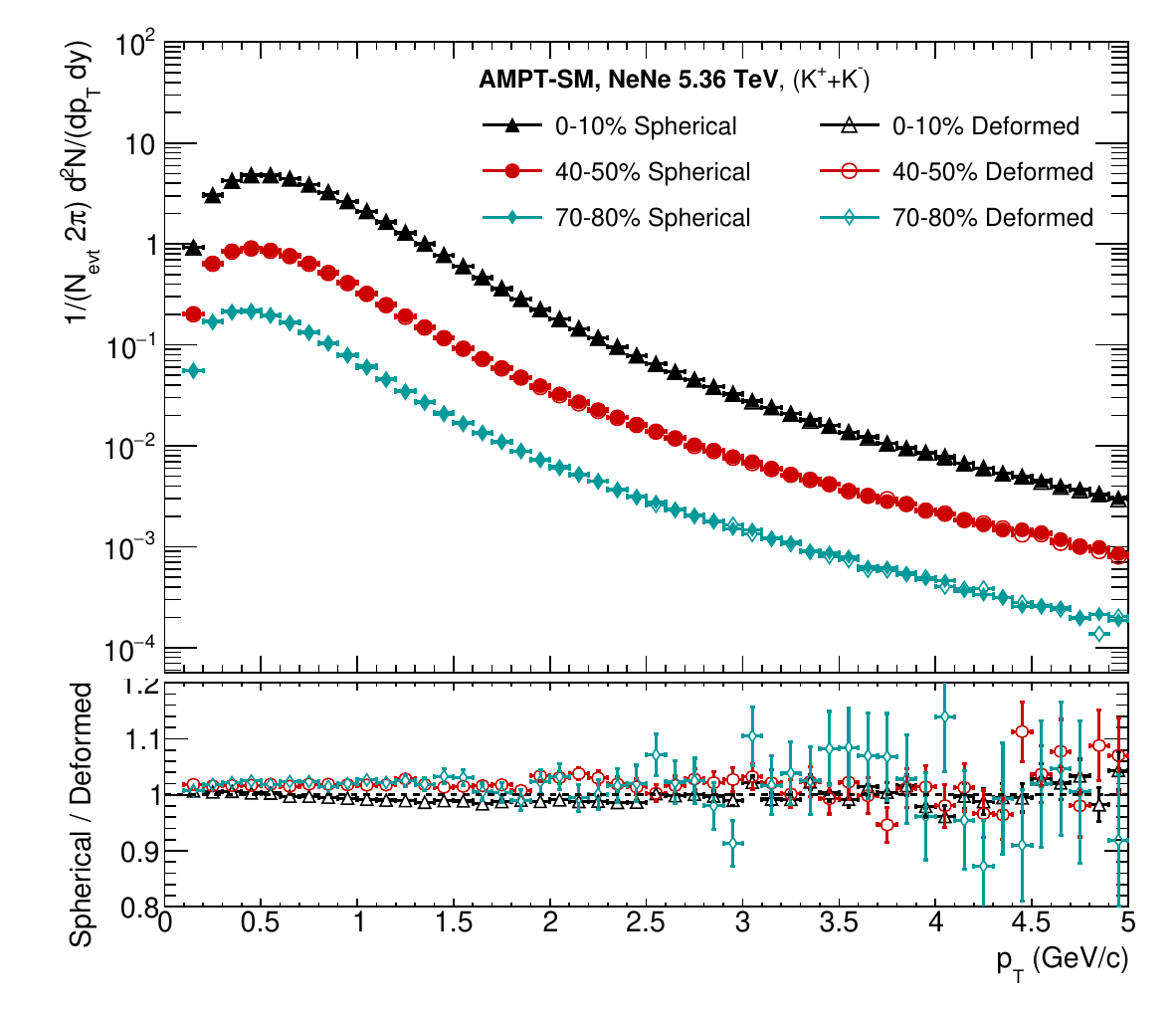}
\end{minipage}\hfill%
\begin{minipage}[t]{0.33\textwidth}
  \centering
  \includegraphics[width=\linewidth]
  {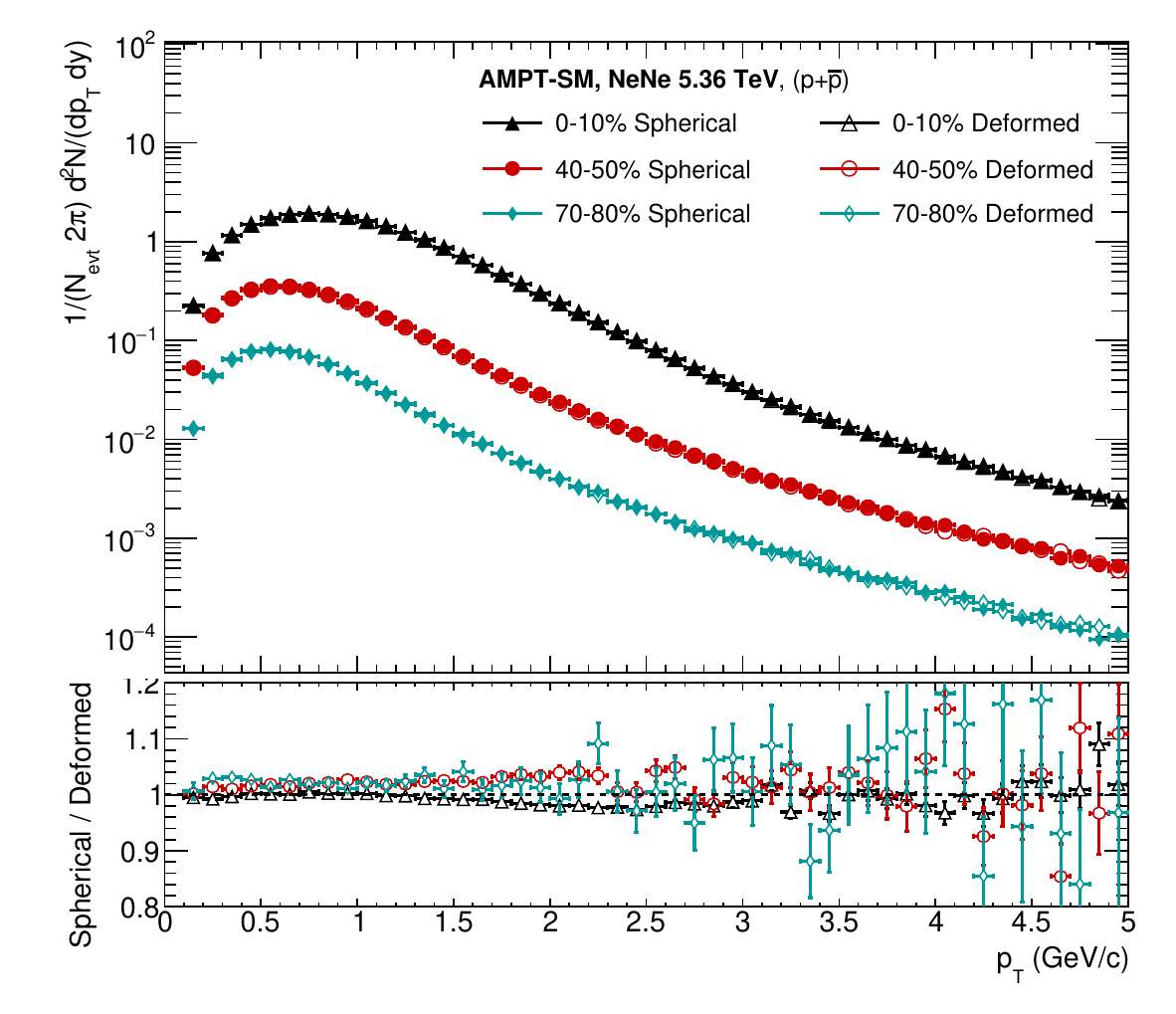}
\end{minipage}

\caption{(Color Online) Transverse momentum (\ppt) spectra of identified hadrons, $\pi^{+}+\pi^{-}$, $K^{+}+K^{-}$, and $p+\bar{p}$, in different centrality intervals for \nene collisions at \five from the AMPT-SM. Results for spherical and deformed configurations of ${}^{20}\mathrm{Ne}$ are shown. The lower panels display the ratio of spherical to deformed spectra. }
\label{fig4}
\end{figure*}
Figure~\ref{fig4} presents the transverse momentum (\ppt) spectra of identified hadrons, $\pi^{+}+\pi^{-}$, $K^{+}+K^{-}$, and $p+\bar{p}$, for different centrality intervals in \nene collisions at \five from the AMPT-SM. The spectra exhibit the expected decrease with increasing \ppt, characteristic of soft particle production. A clear centrality dependence is observed, with larger yields in central collisions compared to peripheral ones, reflecting the increased number of participant nucleons and parton production. A clear mass ordering is observed in the spectral shapes, where heavier particles exhibit harder \ppt distributions compared to lighter ones. In particular, the proton spectra are flatter than those of kaons and pions, indicating a larger average transverse momentum. This behavior is commonly interpreted as a signature of collective radial flow, where a common transverse expansion velocity boosts heavier particles more effectively due to their larger mass. Within the AMPT-SM, such collective behavior arises from the interplay of partonic scatterings and hadronization via quark coalescence. The partonic phase builds up collective motion through multiple scatterings, while the coalescence mechanism translates parton-level flow into hadron-level spectra. Subsequent hadronic rescattering further modifies the momentum distributions, leading to the observed mass-dependent spectral hardening.

A comparison between spherical and deformed configurations of ${}^{20}\mathrm{Ne}$ shows that the overall spectral shapes remain very similar for all particle species and centralities. The ratio of spherical to deformed spectra stays close to unity over most of the \ppt range, indicating that nuclear deformation has a limited impact on the bulk transverse expansion dynamics. This suggests that the collective radial flow developed in the system is primarily driven by the overall energy density and interaction dynamics, rather than the detailed initial nuclear geometry. A modest but systematic deviation from unity is observed in peripheral collisions, where the ratio reaches values at the level of $2$--$6\%$ depending on \ppt and particle species. In these collisions, the particle density is lower and the number of interactions is reduced, allowing a slightly stronger sensitivity to the initial-state geometry to persist. However, even in this regime, the deformation-induced effect remains small compared to the dominant collective dynamics governing the spectral shapes. Overall, these results indicate that while AMPT-SM successfully generates mass ordering and radial flow-like features in the $p_{\mathrm{T}}$ spectra, the influence of nuclear deformation on these bulk observables is weak. The deformation effect appears as a small correction, most visible in peripheral collisions, and does not significantly modify the collective expansion pattern of identified hadrons.

\begin{figure}[h]
\centering
\includegraphics[width=\columnwidth]{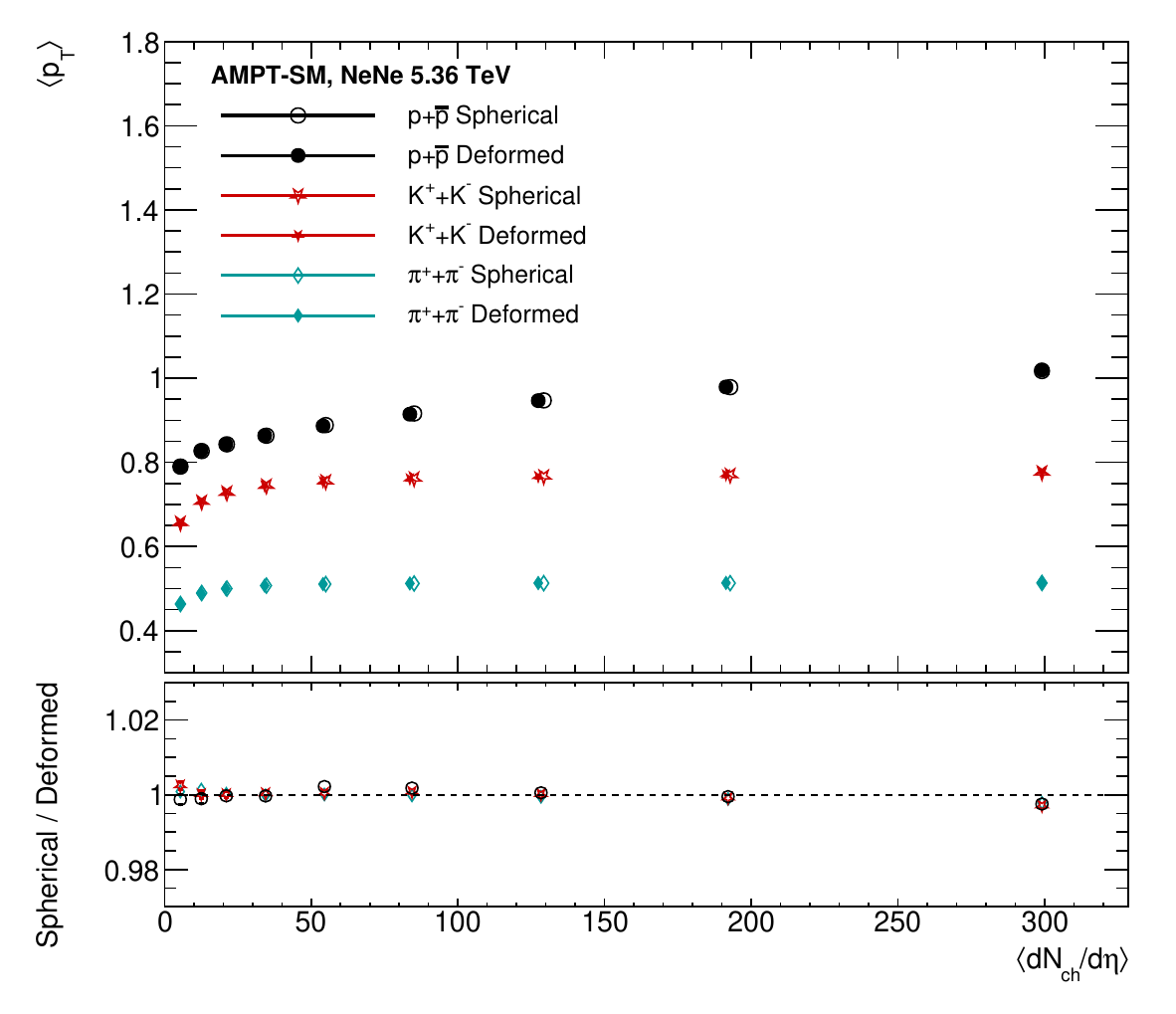}
\caption{(Color Online) Mean transverse momentum $\langle p_{\mathrm{T}} \rangle$, of identified hadrons ($\pi^{+}+\pi^{-}$, $K^{+}+K^{-}$, and $p+\bar{p}$) as a function of the average charged-particle multiplicity $\langle dN_{\mathrm{ch}}/d\eta \rangle$, in \nene collisions at \five from the AMPT-SM. Results for spherical and deformed configurations are shown. The lower panel presents the ratio of spherical to deformed.}
\label{fig5}
\end{figure}

Figure~\ref{fig5} shows the mean transverse momentum ($\langle p_{\mathrm{T}} \rangle$), of identified hadrons as a function of charged particle multiplicity $\langle dN_{\mathrm{ch}}/d\eta \rangle$, in \nene collisions at \five from AMPT-SM. A clear increase of $\langle p_{\mathrm{T}} \rangle$ with multiplicity is observed for all particle species, reflecting the development of stronger transverse collective expansion in higher-multiplicity events. Additionally, the $\langle p_{\mathrm{T}} \rangle$ shows mass ordering with $\langle p_{\mathrm{T}} \rangle_{p} > \langle p_{\mathrm{T}} \rangle_{K} > \langle p_{\mathrm{T}} \rangle_{\pi}$. This behavior is interpreted as a signature of radial flow, where heavier particles receive a larger transverse momentum boost due to the collective expansion of the system. In AMPT-SM, this behavior arises from partonic scattering that build up collective motion, followed by quark coalescence and hadronic scatterings, which transfer the flow to final state hadron. The comparison between spherical and deformed configurations shows that the $\langle p_{\mathrm{T}} \rangle$ values are almost the same over the full multiplicity range. The ratio of spherical to deformed is close to unity within 1\%--2\% deviation indicating that deformation has negligible impact on overall strength of radial flow in \nene collisions within the AMPT-SM. These results demonstrate that collective transverse expansion is primarily governed by the overall system density and interaction dynamics, rather than detailed initial geometry. The small deviations observed, even in peripheral collisions, indicate that deformation effects are subleading and do not significantly modify the bulk evolution of the system.

\begin{figure*}[ht]
    \centering
    \includegraphics[width=0.45\linewidth]{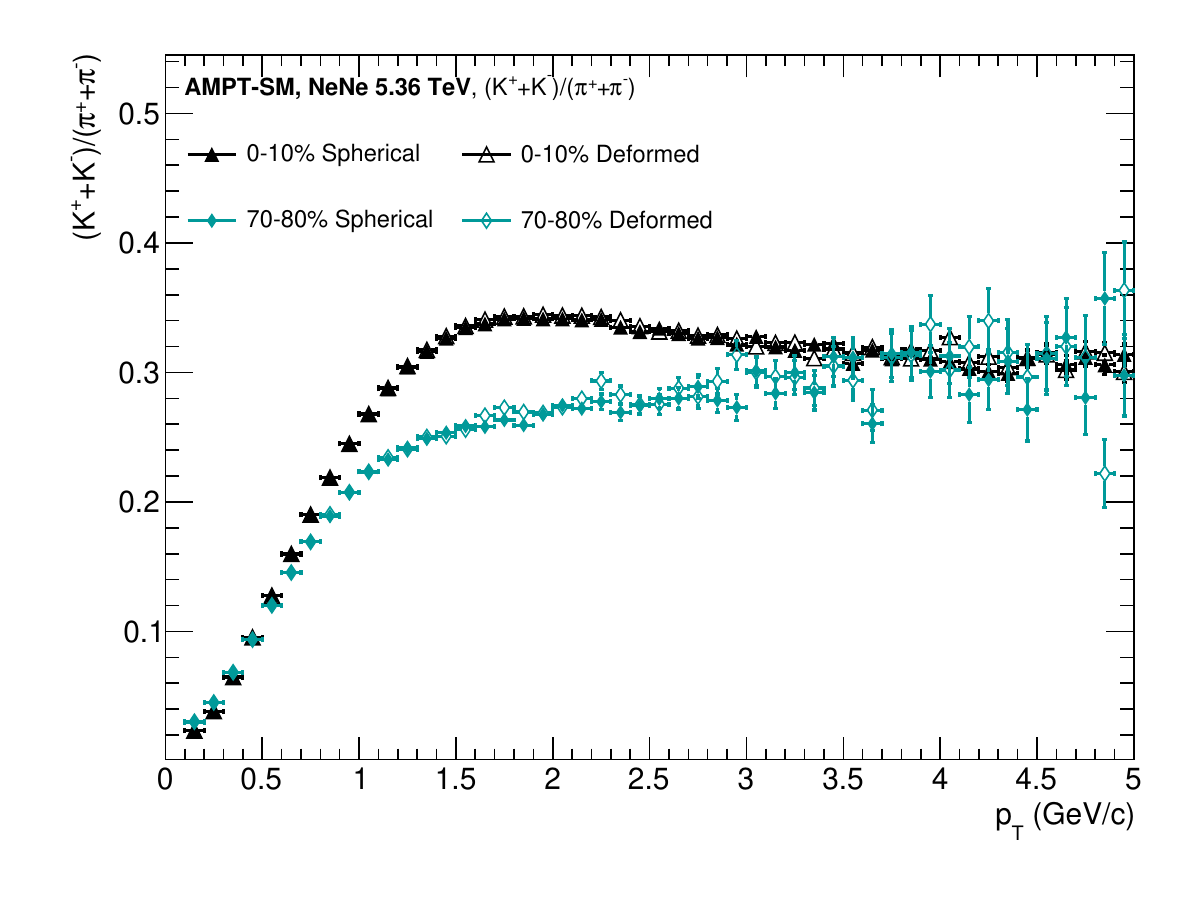}
    \includegraphics[width=0.45\linewidth]{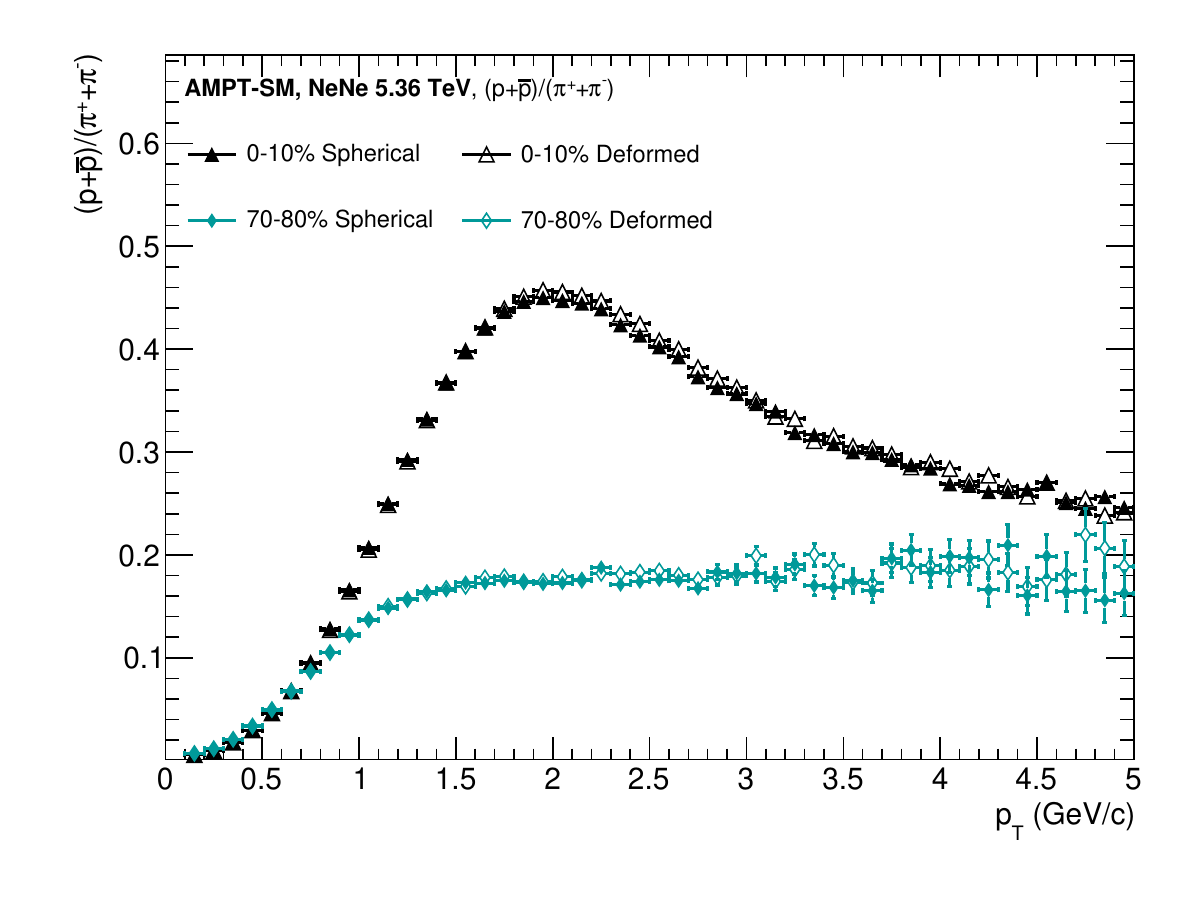}
    \caption{(Color Online) Transverse momentum dependence of particle yield ratios $K/\pi$ and $p/\pi$ for three centrality classes ($0–10\%, 70–80\%)$ in \nene collisions at \five from the AMPT-SM. Results are shown for spherical and deformed nuclear configurations. }
    \label{fig6}
\end{figure*}

Figure~\ref{fig6} shows the transverse momentum dependence of particle ratios $K/\pi$ and $p/\pi$ for 0--10\% central and 70--80\% peripheral \nene collisions at \five from AMPT-SM. These ratios provide hydrochemical composition of the system and are sensitive to the particle production mechanism, including strangeness production and baryon formation. The $K/\pi$ ratio exhibits a smooth increase with \ppt, reflecting the relative enhancement of strange particle production at higher \ppt. On the other hand, the $p/\pi$ ratio shows the expected rise at intermediate \ppt, which is associated with baryon enhancement due to collective radial flow and quark coalescence mechanism. These behaviors are consistent with what has already observed in larger collision systems and indicate that AMPT-SM captures key aspects of hadronization dynamics. The comparison between spherical and deformed configurations shows that both, $K/\pi$ and $p/\pi$ ratios remain similar across full \ppt range and for all centralities. The ratio of the two configurations are close to unity, with small deviation of 1\%--3\% indicating that the nuclear deformation has a negligible effect on the hadrochemical composition in \nene collisions in the AMPT-SM.   

In AMPT-SM, hadronization proceeds via quark coalescence, where particle ratios are determined primarily by the underlying quark phase-space distributions. Since the introduction of nuclear deformation does not significantly modify the overall parton densities or their momentum distributions at fixed multiplicity, the resulting particle ratios remains largely unaffected. This suggests that hadrodynamical observables are primarily geverned by bulk medium properties rather than the detailed initial-state geometry. Overall, these results demonstrate that nuclear deformation has a minimal impact on both strangeness production and baryon-to-meson ratios, reinforcing the conclusion that deformation effects are subleading in bulk hadron production in \nene collisions in AMPT-SM.

\section{Summary and Conclusions}

In this work, we have performed a detailed study of particle production in \nene collisions at \five from the AMPT model with string melting configuration. By comparing spherical and deformed ${}^{20}\mathrm{Ne}$ nuclei, we investigated the sensitivity of bulk observables to initial-state nuclear deformation.

The charged-particle pseudorapidity distributions and midrapidity yields exhibit a smooth centrality dependence and show only minimal differences between the two configurations. Similarly, the rapidity densities of identified hadrons scale with charged-particle multiplicity, indicating that particle production is primarily driven by event activity rather than the detailed initial geometry.

The transverse momentum (\ppt) spectra of pions, kaons, and protons shows the expected centrality dependence and mass ordering, consistent with the presence of collective radial flow. The mean transverse momentum $\langle p_{\mathrm{T}} \rangle$ further confirms this behavior, showing a clear increase with multiplicity and a hierarchy $\langle p_{\mathrm{T}} \rangle_{p} > \langle p_{\mathrm{T}} \rangle_{K} > \langle p_{\mathrm{T}} \rangle_{\pi}$. In all cases, the comparison between spherical and deformed configurations reveals only small differences, typically at the level of $1$--$2\%$, indicating that deformation does not significantly modify the strength of collective expansion.

The particle ratios $K/\pi$ and $p/\pi$ exhibit the expected $p_{\mathrm{T}}$ dependence associated with strangeness production and baryon enhancement. The ratios remain nearly identical for spherical and deformed configurations, with deviations of at most $1$--$3\%$, suggesting that nuclear deformation has a negligible effect on hadrochemical composition.

A slightly enhanced sensitivity to deformation is observed in peripheral collisions, where lower particle density and reduced interaction rates allow a modest imprint of the initial geometry to persist. However, even in this regime, the deformation effect remains subleading compared to the dominant influence of multiplicity and system evolution.

Overall, our results demonstrate that, within the AMPT-SM, nuclear deformation has a limited impact on bulk particle production in \nene collisions. The collective dynamics and hadronization processes largely wash out the differences arising from the initial-state geometry, leading to only minor modifications in final-state observables. These findings provide important guidance for future studies of deformation effects and highlight the need to explore more sensitive observables, such as anisotropic flow, to probe initial geometry in intermediate-size systems.

\section{Acknowledgements}

\bibliographystyle{utphys}
\bibliography{bib}

\end{document}